\documentstyle[emulateapj,apjfonts,psfig]{article}

\lefthead{J.~M.~Miller et al.}  
\righthead{Stellar-mass Black Hole Spins}

\received{Date}
\revised{Date}
\accepted{Date}

\journalid{vol}{date}
\articleid{1}{4}
\paperid{id}

\cpright{AAS}{1999}
\ccc{x}

\begin{document}

\title{Stellar-mass Black Hole Spin Constraints from Disk Reflection
and Continuum Modeling}

\author{J.~M.~Miller\altaffilmark{1},
        C.~S.~Reynolds\altaffilmark{2},
        A.~C.~Fabian\altaffilmark{3},
        G.~Miniutti\altaffilmark{3,4,5},
        L.~C.~Gallo\altaffilmark{6}}

\altaffiltext{1}{Department of Astronomy, University of Michigan, 500
Church Street, Ann Arbor, MI 48109, jonmm@umich.edu}
\altaffiltext{2}{Department of Astronomy, The University of Maryland,
College Park, MD, 20742}
\altaffiltext{3}{Institute of Astronomy, University of Cambridge,
Madingley Road, Cambridge CB3 OHA, UK}
\altaffiltext{4}{Laboratoire APC, UMR 7164, 10 rue A. Domon et
L. Duquet, 75205 Paris, FR}
\altaffiltext{5}{LAEX, Centro de Astrobiolgia (CSIC-INTA), LAEFF,
P.O. Box 78, E--28691, Villanueva de la Ca\~{n}ada, Madrid, ES}
\altaffiltext{6}{Department of Astronomy \& Physics, Saint Mary's
University, 923 Robie Street, Halifax NS B3H 3C3}


\authoremail{jonmm@umich.edu}

\label{firstpage}

\begin{abstract}
Accretion disk reflection spectra, including broad iron emission
lines, bear the imprints of the strong Doppler shifts and
gravitational red-shifts close to black holes.  The extremity of these
shifts depends on the proximity of the innermost stable circular orbit
to the black hole, and that orbit is determined by the black hole spin
parameter.  Modeling relativistic spectral features, then, gives a
means of estimating black hole spin.  We report on the results of fits
made to archival X-ray spectra of stellar-mass black holes and black
hole candidates, selected for strong disk reflection features.
Following recent work, these spectra were fit with reflection models
and disk continuum emission models (where required) in which black
hole spin is a free parameter.  Although our results must be regarded
as preliminary, we find evidence for a broad range of black hole spin
parameters in our sample.  The black holes with the most relativistic
radio jets are found to have high spin parameters, though jets are
observed in a black hole with a low spin parameter.  For those sources
with constrained binary system parameters, we examine the distribution
of spin parameters versus black hole mass, binary mass ratio, and
orbital period. We discuss the results within the context of black
hole creation events, relativistic jet production, and efforts to
probe the innermost relativistic regime around black holes.

\end{abstract}

\section{Introduction}
X-rays probe the innermost regime around compact objects.  Modern data
aside, this must hold true based on simple theoretical considerations.
Temperature relations for standard blackbody accretion disks and the
energy budget of scattered emission that might account for hard X-ray
production both demand that X-ray emission arise close to compact
objects.  Thus, the promise of X-ray studies of black holes is that it
may allow observers to study the innermost regime, and so explore
general relativity in the strong field limit.  The promise is
especially rich in the case of stellar-mass black holes, because the disk
is itself an X-ray object, and spectra are not complicated by a
stellar surface and boundary layer as in the case of neutron stars.

Measuring spin in stellar-mass black holes can do more than confirm
the predictions of the Kerr metric.  Unlike supermassive black holes
in galactic centers, stellar-mass black holes likely gain most of
their angular momentum at the moment of their birth (Volonteri et al.\
2005, Gammie et al.\ 2004).  Spin is therefore a window into the
nature of the supernovae and/or gamma-ray bursts that give rise to
stellar-mass black holes (many progenitor star properties are
important; see Heger \& Woosley 2002).  Some models for jet production
in black holes rely on tapping the spin energy of the hole, and so
predict a link between spin and jets (e.g. Blandford \& Znajek 1977).
In the case of AGN, radio loudness is often used as a proxy for spin
(see, e.g., Sikora, Stawarz,\& Lasota 2007); this proxy creates a
circularity in efforts to understand jet production mechanisms.  While
the sample of AGN is much larger than the sample of stellar-mass black
holes in the Milky Way, the proximity of stellar-mass black holes
facilitates direct investigations of spin that can bear on jet
mechanisms.

Iron emission lines formed in the inner accretion disk will bear the
imprints of the strong Doppler shifts and gravitational red-shifts
endemic to that region, and so can serve as incisive probes of the
innermost relativistic regime (for a review, see Miller 2007).  A
major advantage of disk lines is that the mass of the black hole is
not required to make a spin measurement, and other parameters -- such
as the inner disk inclination -- can be measured directly.  In recent
years, these lines (and the larger disk reflection spectrum, of which
iron lines are the most prominent part) have been used to obtain
general constraints on black hole spin.  (The dimensionless spin
parameter is given by $a = cJ/GM^{2}$ and values range from zero to
one).  For instance, line profiles in XTE J1650$-$500 and GX 339$-$4
could be fit well with a model appropriate for a disk around a
maximally-spinning black hole (Miller et al.\ 2002, Miniutti, Fabian,
\& Miller 2004, Miller et al.\ 2004a, 2006; Laor 1991).

It is only within the last two years that fits to spectra have started
to yield non-zero spin constraints that could properly be called
measurements.  A spin of $a>0.98$ has been reported for the Seyfert-1
AGN MCG-6-30-15 (Brenneman \& Reynolds 2007), and a spin of $a =
0.93(5)$ has been reported in GX 339-4 (Miller et al.\ 2008; see also
Reis et al.\ 2008).  This advance is partly due to improved spectra
and partly due to the development of new, variable-spin line models.
It is now clear that relativistic disk lines are also present in
neutron star spectra (see, e.g., Bhattacharyya \& Strohmayer 2008,
Cackett et al.\ 2008), providing a means of constraining fundamental
neutron star parameters and a useful comparison for black hole
spectra.

The accretion disk continuum can, in principle, also be exploited to
constrain black hole spin.  Especially in the case of stellar-mass
black holes, wherein the disk is an X-ray object and can 
dominate the total spectrum, this method is promising.  The
development of new disk models, supported by numerical disk
simulations, has enabled spin constraints in a few systems based on
the thermal continuum emission from the disk (Shafee et al.\ 2006,
McClintock et al.\ 2006).  This method essentially
amounts to measuring the emitting area of the accretion disk, and so
requires accurate knowledge of a given black hole mass, its
distance, and a detector with an accurate flux calibration.  

Quasi-periodic oscillations (QPOs) in the X-ray flux of accreting
compact objects may provide an incisive way of studying orbital motion
near to black holes (for a review, see van der Klis 2006).  A number
of difficulties persist, however.  The high frequency (100 Hz and
above) QPOs that are most plausibly associated with inner orbits
around black holes are too hard (spectrally) to arise via direct
emission from the accretion disk (see, e.g., Homan et al.\ 2001).
Moreover, shifts in QPO frequencies are not easily interpreted as due
to changes in the inner edge of the disk as they can occur at high
mass accretion rates where the disk must be at its innermost stable
circular orbit (ISCO).  Frequencies and frequency changes may have a
complex dependence on a combination of the inner disk radius, mass
accretion rate, and other paramters (see, e.g., van der Klis 2001).
At present, there is no theory that can fully explain the frequencies
observed, how such frequencies might arise, and their energy budget
(for a discussion, see Reynolds \& Miller 2008).

It is now possible to make a systematic spectral analysis of spin in
stellar-mass black holes that exploits all of the best, most
physically-motivated models.  We have identified a set of eight
stellar-mass black holes and black hole candidates, selected for
having strong, broad iron lines.  Whenever possible, data from CCD
spectrometers was used in order to take advantage of moderate spectral
resolution.  In order to construct the most self-consistent models
possible, we fit all of the spectra with blurred disk reflection
models and disk continuum models in which spin is a variable.  In each
case, the spin parameter in the disk continuum and blurred disk
reflection models were linked, in order to obtain the most robust
constraints possible.  In the sections that follow, we detail aspects
of the source sample, our analysis methods, the results of our
analysis, sources of systematic errors and biases, and possible
implications.

\section{Observations and Data Reduction}
Optical and infrared observations have constrained the properties of
six of the eight binary systems in our sample.  The parameters of
these systems are reported in Table 1.  In some cases, such as
GRO~J1655$-$40, values have been measured precisely; however,
this is not typical.  The numbers quoted in Table 1 reflect our best
estimates of the value and uncertainty in each parameter of interest.
Where uncertainties are a large fraction of the value itself, a range
is given instead of a value and associated error. 

\subsection{4U 1543$-$475}
We analyzed the very high state spectrum with the largest line
equivalent width, as measured by Park et al.\ (2004).  This
observation was made on 2002 July 19, and is archived as
70133-01-29-00.  The {\it RXTE} standard products from the archive
were utilized.  These files include source and background spectra as
well as instrument responses.  The standard PCA source spectrum is a
sum of all layers from PCU-2 and PCU-3.  The HEXTE cluster-B spectrum
was analyzed jointly with the PCA spectrum.  The net exposures were
1.1 ksec and 0.4 ksec for the PCA and HEXTE, respectively.

We followed standard practices in fitting the PCA and HEXTE spectra.
We added 0.6\% systematic errors to the PCA spectrum using the ftool
``grppha''.  Within XSPEC, the PCA spectrum was fit in the
2.8--25.0~keV range, and the HEXTE spectrum was fit in the 20.0--50.0
keV range.  The upper limit for the HEXTE spectrum is the highest
energy at which the source is confidently detected.  All fitting
parameters were linked between the PCA and HEXTE spectra; an overall
normalizing constant allowed to float between the spectra.  

\subsection{XTE J1550$-$564}
The best available disk reflection spectra of XTE J1550$-$564 were
obtained with {\it ASCA} during the 1998 outburst.  We analyzed the
{\it ASCA}/GIS spectra previously discussed in Miller et al.\ (2005).
These spectra were accumulated on 1998 September 23 in the very high
state; a net exposure of 25 ksec was obtained.  The standard source
and background files and responses from the {\it ASCA} standard
products (available through HEASARC) were analyzed.  The reduction of
these data follows the method best suited to bright sources, outlined
by Brandt et al.\ (1996).

We fit the GIS-2 and GIS-3 spectra jointly in the 1--10~keV band.  The
spin parameter, inner disk inclination, reflection fraction, and
ionization parameter were linked in joint fits to the spectra, and
linked between model components where appropriate.  Other parameters
relevant to the continuum were allowed to float.  This is a pragmatic
approach driven by the fact that different cameras, even on the same
observatory, never have perfectly consistent flux calibrations.  For
clarity and simplicity, the parameters measured from GIS-2 are
reported in Table 2 and Table 3.

\subsection{XTE J1650$-$500}
At present, the mass of this black hole has not been precisely
determined, but preliminary constraints have been made (see Orosz et
al.\ 2004).  In this work, we assume a mass range 3--7~$M_{\odot}$ for
the black hole.  The upper mass limit is based on an assumed mass
ratio of $q\simeq 10$ (Orosz et al.\ 2004).  A mass ratio of $q = 10$
is therefore assumed in this work, and we attach fiducial errors of
30\% to be conservative.  XTE J1650$-$500 was observed with {\it
XMM-Newton} on 2001 September 13.  We analyzed the same EPIC-pn
spectrum discussed in Miller et al.\ (2002).  The spectrum was
accumulated over 21 ksec.  The spectral parameters reported by Miller
et al.\ (2002) and Miniutti, Fabian, \& Miller (2004), especially when
viewed in the context of the extensive work done by Rossi et al.\
(2005), suggest the source was in a rising phase of the low/hard state
when observed with {\it XMM-Newton}.

Owing to calibration uncertainties in pn ``burst'' timing mode, we
restricted our fits to the 0.7--10.0~keV range.  This range was also
used in recent fits to the EPIC-pn ``burst'' mode spectrum of GX
339$-$4 in the very high state (Miller et al.\ 2004, 2008).  A
Gaussian with zero width was included at 2.3~keV to account for
calibration uncertainties in that range.  Imperfect modeling of
silicon features and gold features continue to have small effects in
the 2--3 keV range in ``burst'' mode.  These narrow-band calibration
problems do not affect results related to the continuum, iron line, or
reflection.

\subsection{GRO J1655$-$40}
Although the fundamental parameters of this binary are known
precisely, GRO J1655$-$40 is a challenging source.  The rich
absorption spectrum observed in some very high and high/soft states
of GRO J1655$-$40 (e.g. Miller et al.\ 2006b, Diaz Trigo et al.\ 2007,
Miller et al.\ 2008b) can serve to complicate disk reflection
spectroscopy.  Absorption lines from Fe XXV and Fe XXVI fall in the
midst of the relativistic Fe disk line.  Prior claims for strong
Doppler-shifted emission lines by Balucinska-Church and Church (2000),
for instance, might be partially explained by an absorption line
amidst a broad emission line.

A spectrum with a strong power-law component (for constrast) and
little absorption is optimal for spin constraints.  Archival spectra
of bright states obtained with {\it ASCA}, {\it Chandra}, and {\it
XMM-Newton} are all complicated by absorption.  A recent observation
of GRO J1655$-$40 in the low/hard state with {\it Suzaku} does not
find an Fe disk line (Takahashi et al.\ 2008); improved calibration
may enable more detailed studies in the future.  Thus, although a
focusing telescope and CCD resolution are strongly preferred, we
turned to {\it RXTE}.  The large number of observations that {\it
RXTE} executes makes it possible to find a more suitable spectrum.

{\it RXTE} observed GRO J1655$-$40 during its outburst in 1996.
Sobczak et al.\ (1999) made fits to these spectra.  We selected the
very high state observation with the strongest power-law flux, for
best overall contrast.  Sobczak et al.\ (1999) do not find evidence
for absorption in these spectra, and the power-law index was found to
be reasonably hard ($\Gamma = 2.64$).  

As with 4U 1543$-$475, the standard products (source and background
files, and response files) from the public archive were analyzed.  The
spectra were obtained on 1996 November 2; net exposures of 4.9 ksec
and 1.6 ksec were recorded with the PCA and HEXTE, respectively.  We
added 0.6\% errors to the PCA spectrum using ``grppha''.  All fit
parameters between the PCA and HEXTE spectra were linked, apart from
an overall normalizing constant that was allowed to float between
them.  

\subsection{GX 339$-$4}
GX 339$-$4 is a recurrent Galactic black hole binary that has
undergone numerous outbursts.  At the time of writing, it is the only
black hole that has been observed in {\it every} outburst state with
CCD and/or dispersive X-ray spectrometers.  While the orbital period
of GX 339$-$4 is well known, the mass of the companion star is not.
Hynes et al.\ (2003) and Munoz-Darias et al.\ (2008) have
independently tried to estimate the mass of the black hole in GX
339$-$4.  For the purposes of this analysis, we adopt conservative
values from these analyses: $M_{BH} \geq 6~M_{\odot}$ and $q \geq 8$.

It is difficult to obtain the black hole mass precisely because light
from the accretion disk ``contaminates'' light from the companion even
in quiescent phases in GX 339$-$4, complicating measurements of its
inclination.  Constraints derived from jet flux ratios in this system
strongly suggest a low inclination for the inner disk (e.g., $\theta
\leq 30^{\circ}$, see Gallo et al.\ 2004 and Miller et al.\ 2004).
The inclination of the inner disk may be lower than the inclination of
the binary.  The inclination was a free parameter in all fits reported
in this work.

We fit the same very high state spectrum detailed in Miller et al.\
(2004) and Miller et al.\ (2008).  This EPIC-pn ``burst'' mode
spectrum was obtained on 2002 September 29; a 75.6~ksec exposure was
recorded.  Prior analysis of the disk line and disk reflection
spectrum in GX~339$-$4 suggests a black hole spin parameter of $a =
0.93 \pm 0.05$ (Miller et al.\ 2008, Reis et al.\ 2008).  These
measurements draw on multiple excellent spectra and appear to be
independent of the disk reflection model used; they are likely to be
fairly robust.  However, prior modeling efforts made use of more
phenomenological models for the thermal disk continuum.  While the
more physical disk continuum model used in this work does not give a
statistically superior description of the disk spectrum in GX 339$-$4,
it is more physical.  As with other ``burst'' mode spectra, we fit the
spectrum of GX 339$-$4 over the 0.7--10.0~keV range.

\subsection{SAX J1711.6$-$3808}
SAX J1711.6$-$3808 was observed in outburst as a moderately-bright
transient in 2001.  The parameters of the binary system have not yet
been constrained.  Observations made by {\it BeppoSAX} and {\it RXTE}
are reported by in 't Zand (2002) and Wijnands \& Miller (2002).  The
source was observed by {\it XMM-Newton} and a relativistic iron line
has been reported (Sanchez-Fernandez et al.\ 202, AN, 327, 1004), but
the spectrum is complicated by photon pile-up and so excluded from our
analysis.

We analyzed the {\it BeppoSAX} MECS spectrum previously examined by in
't Zand et al.\ (2002).  The source was observed with the MECS starting
on 2001 February 16.  A net exposure of 37 ksec was obtained.  We fit
the single MECS spectrum in the 2-10 keV range.  Thermal disk emission
was not detected in this spectrum.  This is likely due in part to the
relatively high line of sight absorption ($N_{H} = 2.57(7) \times
10^{22}~ {\rm cm}^{-2}$), which would be especially effective in
hiding a cool disk (see, e.g., Miller et al.\ 2006).  The hard
power-law index and lack of a hot disk component strongly suggest that
SAX J1711.6$-$3808 was observed in the low/hard state.  Parameters
such as the black hole mass and distance are not required in disk
reflection fits; the inner disk inclination was allowed to vary
freely.

\subsection{XTE J1908$+$094}
Like SAX J1711.6$-$3808, XTE J1908$+$094 was observed as a moderately
bright transient with characteristics typical of black hole systems.
At present, the parameters of this binary system are unknown.

We analyzed the {\it BeppoSAX} MECS spectra previously examined by in
't Zand et al.\ (2002b).  The source was observed starting on 2002
April 2 for a total of 56.6 ksec.  XTE J1908$+$094 was found to be
highly variable during this time, so in 't Zand et al.\ (2002b)
examined seven spectra from different time slices.  To facilitate
comparisons with this prior work, we followed the same procedure.  The
same spectra were fit jointly in the 2--10 keV band.  Quantities such
as power-law index and flux were allowed to float between
observations, but the disk reflection parameters and spin parameters
were linked.  As with SAX J1711.6$-$3808, thermal emission from the
disk was not detected, likely due to a combination of high
line-of-sight absorption and the expectation of an intrinsically cool
disk in the low/hard state.  Here again, the current lack of
constraints on the system parameters poses no difficulty for disk
reflection modeling.  The inclination and other parameters were
allowed to vary freely in all fits.

\subsection{Cygnus X-1}
Efforts to measure the parameters of this binary system are
complicated by the fact that the companion is an O9.7 Iab supergiant,
and by the fact that the system is persistently active.  As the
inclination is poorly known, this parameter was allowed to float
freely in all spectral fits.

We analyzed two {\it XMM-Newton}/EPIC-pn spectra of Cygnus X-1 in its
``high'' state.  It should be noted that the ``high'' state in Cygnus
X-1 is an archaic label; the same phase is called the ``very high''
state in other sources.  These spectra were obtained in ``burst''
timing mode, like the spectrum of XTE J1650$-$500.  Preliminary fits
to one of these spectra are discussed in Miller et al.\ (2007).  The
spectra were obtained on 2004 October 6 and 2004 October 8; a net
exposure of 0.5 ksec were obtained in each case.  This is relatively
short; however, the proximity of Cygnus X-1 and its high flux state
when observed yielded sensitive spectra.  Events were extracted using
a narrow strip along the full height of the DETX-DETY plane.  Spectral
files were produced by selecting event grades 0--4 and
enforcing ``FLAG=0'' within ``xmmselect''.  Spectral channels
0--20,479 were grouped by 5 as required for pn analysis.  The
``rmfgen'' and ``arfgen'' tools were used to create response files.
As with other ``burst'' mode spectra, we fit the spectra of Cygnus X-1
on the 0.7--10.0~keV band.  The spectra were fit jointly.  The spin
parameters were linked, but as the spectra were not simultaneous all
other parameters were allowed to vary independently.  For clarity and
convenience, the result of fits to the first observation are listed in
Table 2 and Table 3.

In the fits described below, we allowed for a narrow Gaussian
absorption line corresponding to Fe XXV or Fe XXVI to account for any
absorption in the massive companion wind.  The width of the Gaussian
was fixed to zero, as wind velocities are below the resolution of the
detector.  The addition of such a line gave only marginal
improvements, and equivalent widths of approximately 30 eV and 10 eV
for the first and second observations, respectively.

\subsection{Conspicuous Absentees}
GRS 1915$+$105 is an important but complex source requiring
individual treatment, and so it has been left out of this analysis.
Prior fits to the disk line in this source did not require black hole
spin; indeed, the fits were suggestive of a low spin parameter
(Martocchia et al.\ 2002).  More recent fits to the thermal continuum
have suggested a near-maximal spin (McClintock et al.\ 2006).  More
work is needed to resolve this disparity.  The results of disk
reflection modeling of a recent {\it Suzaku} spectrum of GRS
1915$+$105 will be reported by Blum et al.\ (2009).

Iron emission lines are evident in the spectra of V4641 (in 't Zand
2000), and those lines can indeed be modeled as relativistic disk
lines (Miller et al.\ 2002b).  However, it is not clear that iron
lines can be used to study the inner accretion flow in this source.
Emission lines from the inner disk may be contaminated by lines from a
surrounding nebula or outflow.  This possibility was first recognized
by Revnivtsev et al.\ (2002) based on observations made in outburst.
More recent {\it Chandra} observations of the source in a nearly
quiescent flux state also reveal iron emission lines (Gallo et al.\
2009).

\section{Analysis and Results}
\subsection{Models and Methodology}
Using the inner disk as an indirect measure of the black hole spin
parameter depends on two assumptions; firstly, that the accretion flow
has a sharp transition from turbulent orbital flow to an inward
plunging flow at (or close to) the innermost stable circular orbit;
and second that the iron line emission has an inner truncation radius
due to this transition in the flow properties.  Recent 3-dimensional
magnetohydrodynamic simulations of black hole disks suggest that these
assumptions are valid if the disk is sufficiently thin (Reynolds \&
Fabian 2008; Shafee et al. 2008).  This condition likely holds for the
range of outburst phases covered in this work, including relatively low
accretion rates in the low/hard state (see Miller et al. 2006b; Rykoff
et al. 2007).  Therefore, in selecting spectra for this analysis,
bright phases of the low/hard state are taken to be as relevant and
well-suited as observations made in brighter states.

All spectra were fit using XSPEC version 11.3 (Arnaud \& Dorman 2000).
The spectra from each source were fit with a model consisting of the
``kerrbb'' thermal disk continuum model (Li et al.\ 2005; this is the
same model employed by Shafee et al.\ 2006 and McClintock et al.\
2006) and the constant density ionized (CDID) disk reflection model
(Ballantyne, Ross, \& Fabian 2001).  This reflection model is
calculated in the fluid frame.  To translate this spectrum into that
seen by a distant observer, we convolved the CDID model with the
``kerrconv'' model (Brenneman \& Reynolds 2006).  ``Kerrconv'' encodes
the Doppler and gravitational shifts expected close to a black hole,
as a function of the black hole spin parameter.  Convolving a
reflection spectrum is more physically self-consistent than merely
treating the relativistic line, because the line and broad-band
reflection spectrum are produced in the same physical location through
the same process.

The black hole spin parameter $a$ and inner disk inclination $\theta$
are common to the ``kerrbb'' and ``kerrconv'' models.  These
parameters were linked in fits to each spectrum of a given source, and
linked in spectra from different observations of the same source.
That is, the disk reflection and continuum spectra {\it jointly}
determined the spin parameter and inner disk inclination in our fits.
For the two sources where thermal emission from the disk was not
detected, spin constraints were obtained through the disk reflection
spectrum alone.

Apart from the black hole spin parameter and the inner disk
inclination, important parameters in the ``kerrconv'' model inlcude the
inner disk radius (in units of the ISCO) and the reflection emissivity
index.  The inner radius was fixed at 1.0; this amounts to assuming
that the disk is at the ISCO. The emissivity is taken to be a
power-law in radius of the form $J(r) \propto r^{-q}$.  A simple
lamp-post model gives $q=3$.  While prior work assumed this form
partially for simplicity, recent microlensing observations provide an
interesting physical justification (Chartas et al.\ 2008).  Other
results have found steeper emissivity indices, perhaps suggesting that
the hard component is anisotropic (Miniutti \& Fabian 2004).  This may
be consistent with hard X-ray emission in the base of a jet.  Whatever
the proper physical picture, this parameter is one that can be
directly constrained by data.  In all fits, the emissivity index was
constrained to be in the range $3 \leq q \leq 5$; this range is
commensurate with that found in the literature (see, e.g., Miller et
al.\ 2007).

The CDID disk reflection model is capable of handling a broad range of
disk ionization, including high ionization states.  The ionization
parameter is used to express a ratio of flux to gas density: $\xi =
{\rm L}_{\rm X} / {\rm n} {\rm r}^{2}$, where $n$ is the hydrogen
number density.  This is important in X-ray binaries; reflection
occurs in the upper levels of the disk, not the midplane, and that
region is expected to be highly ionized.  Results reported in the
literature vary between ${\rm log}(\xi) = 3$ and ${\rm log}(\xi) = 5$,
depending on the source state and luminosity (see Miller 2007).  This
range defines the bounds adopted in our fits.  While this model does
not include thermal emission from the disk midplane, it appears that
neglecting this does not markedly change the spin parameter obtained (see
Miller et al.\ 2008 and Reis et al.\ 2008).

The CDID model can be applied as a ``pure'' reflection model, with no
hard power-law flux included, but we employed a version that includes
a power-law continuum flux.  Important parameters in the CDID model
that must be constrained include: the photon power-law index (the
steepest power-law index possible is $\Gamma = 3$), the disk
ionization parameter, the disk reflection fraction ($0 \leq R \leq
2$), and an overall normalizing factor.  The normaliztion does not
account for distance dilution and is a small number.  The CDID model
is an angle-averaged model: the inner disk inclination is not a
variable that can be constrained by fitting.  The version of the model
that we used assumes solar abunances for all elements.  To obtain good
fits at the limits of a bandpass when using a convolved
reflection model, it is necessary to extend the range over which the
model is calculated within XSPEC using the ``extend'' command.

The ``kerrbb'' model has 10 parameters, including the black hole spin
parameter and the inner disk inclination (Li et al.\ 2005).  The full
parameter list includes: the ratio of power due to torque at the ISCO
to power arising from accretion (fixed to zero in all fits as per a
Keplerian disk); the mass of the black hole in units of solar masses
(constrained to lie in the ranges indicated in Table 1); the mass
accretion rate through the disk (allowed to float in all fits); the
distance to the black hole in units of kpc (constrained to lie in the
ranges indicated in Table 1); the spectral hardening factor, a
multiplicative factor accounting for radiative transfer through the
disk atmosphere (the default value of 1.7 was assumed in all fits); a
switch controlling self-irradiation (all fits neglected
self-irradiation as per the default); a switch controlling limb
darkening (turned off in all fits as per the default); and the overall
model normalization.  If all of the system parameters were known
precisely, the normalization should be set to unity.  In practice, the
normalization serves to absorb difficulties with the system
parameters, the flux calibration of the detector used, and other
effects (see, e.g., Zimmerman et al.\ 2005).

To understand the influence of the disk continuum in estimating spin
parameters, we also made independent fits with the ``diskbb'' model
(Mitsuda et al.\ 1984) taking the place of the ``kerrbb'' model.  This
model has only two parameters, the color temperature of the inner disk
and a flux normalization parameter.  Three aspects of this model make
it powerful: it is simple, and thus easily reproducible; it fits
thermal spectra extremely well; and it has a long history in the
literature, facilitating useful comparisons.  Important physics, such
as a zero-torque inner boundary condition and the effects of radiative
transfer effects, are not included in the ``diskbb'' model.

\subsection{Spectral Fitting Results}
The results of our spectral fits are detailed in Table 2 and Table 3.
Figures 1--16 depict fits to the spectra of each source with a simple
continuum model (in order to highlight the disk reflection features),
and fits made with the fully relativistic, physical models from which
spin parameters are inferred (those described in Table 2).  A
histogram of the spin values listed in Table 2 is shown in Figure 17.
The fits are imperfect; in many cases, they are not formally
acceptable in a statistical sense.  Where fits are not formally
acceptable, the fit statistic is driven to unacceptable values mostly
by instrumental features (see, e.g., Miller et al.\ 2004, for a
discussion of features in the response of the {\it XMM-Newton}/EPIC-pn
camera when operated in burst mode).  It is clear in Figures 1--16
that the relativistic, physical models account for the disk reflection
features very well.

Fits to the spectra of GRO J1655$-$40 and GX 339$-$4 suggest spin
parameters of $a = 0.98(1)$ and $a =0.94(2)$ respectively ($1\sigma$
statistical errors).  XTE J1650$-$500 and XTE J1550$-$564 may also
have high spins, with values of $a = 0.79(1)$ and $a = 0.76(1)$,
respectively. In each of these four cases, relatively high spin values
might have been anticipated based on prior suggestions, whether from
various interpretations of high-frequency QPOs, previous fits to the
Fe K line shape, or extreme jet phenomena.  It should be noted that
the implied inclination of GX 339$-$4 does not imply a very high black
hole mass based on the work of Hynes et al.\ (2003), as the
inclination of the inner disk and binary system need not be the same
(see Maccarone 2002).

The spin obtained for XTE J1908$+$094 is nominally rather high as
well, at $a=0.75(9)$.  The value obtained for SAX J1711.6$-$3808 is
more moderate, at $a=0.6(2)$.  These two values are discussed in more
detail below.
				   
Within our sample, Cygnus X-1 stands out as the only high-mass X-ray
binary.  The observations of Cygnus X-1 considered in this work were
made in a ``high'' state, where the Fe K line is known to be stronger,
perhaps broader (Cui et al.\ 1998), and perhaps more suggestive of
emission from the ISCO (Gilfanov, Churazov, \& Revnivtsev 2000) than
in the ``low/hard'' state.  At $a=0.05(1)$, Cygnus X-1 is ostensibly
found to harbor a black hole with very low spin.  It is interesting to
note that fits to the black hole with the next highest companion mass,
4U 1543$-$475, also give a low spin parameter of 0.3(1). 

To better understand possible physical connections, Figures 18--20
plot the spin values that we have obtained versus black hole mass,
versus the ratio of black hole mass to companion mass, and versus
orbital period using the parameters in Table 1 and Table 2.  There is
no clear evidence that spin is correlated with any of these
parameters.  The most promising case for a correlation is between the
black hole spin parameter and binary mass ratio (see Figure 19).
However, more sources and improved errors on the ratio values needed
to make any relation significant.

\subsection{On the Robustness of the Spin Constraints}
When fitting a complex model in packages such as XSPEC or ISIS,
special care is needed.  Convolution models, table models, and total
models with many individual components are especially prone to saddle
points in $\chi^{2}$ space.  In these cases, efforts to minimize
$\chi^{2}$ are improved with a hands-on approach.  In our analysis,
fits were made within XSPEC in the usual way -- until the minimum
required change in $\chi^{2}$ had nominally been met.  Then, we pushed
random parameters off of their best-fit values by factors of 10--30\%,
and re-fit the data.  This procedure was repeated many times in an
effort to make sure that a {\it global} minimum was found.  Then, we
calculated $1\sigma$ errors on the parameters in the model using the
``error'' command, which calculates a joint error in that it allows
other parameters to float.  

Although this fitting procedure likely improved the rigor of our
results, the errors quoted in Table 2 and Table 3 do not fully convey
the nature of some spin values.  In some cases, the $\chi^{2}$ space
is complex, and though there is a nominal preference for a particular
spin value, extremal values may not excluded at high significance.  We
examined this possibility using the ``steppar'' command within XSPEC.
This command affords greater control over how the $\chi^{2}$ space is
searched.  For each source, the spin parameter was frozen at twenty
evenly-spaced values between zero and unity while all other parameters
were allowed to vary.  We checked the results of steppar by manually
executing the same procedure.  The manual procedure gave more
conservative results than steppar for GRO J1655$-$40; the conservative
results are plotted in Figure 21.  Figures 21--23 plot the dependence
of $\chi^{2}$ on the black hole spin parameter $a$, for each black
hole in our sample.

The spin parameters measured for XTE J1550$-$564 and GX~339$-$4
exclude $a=0$ at far more than the 8$\sigma$ level of confidence; for
GRO J1655$-$40 and XTE J1650$-$500, zero spin is excluded at the
6$\sigma$ level of confidence (see Figure 21).  The results from
Cygnus X-1 strongly suggest a low spin value, and we performed a
similar check on that result.  High spin values (e.g. $a>0.9$) for
Cygnus X-1 at more than the 8$\sigma$ level of confidence (see Figure
22).  However, other spin constraints are less certain.  For 4U
1543$-$475, $a=0$ is only excluded at the $3\sigma$ level of
confidence, although high spin is excluded at more than the 8$\sigma$
level of confidence.  The spins parameters reported for SAX
J1711.6$-$3808 and XTE J1908$+$094 are not confidently determined (see
Figure 23).  Zero spin is just outside of the $1\sigma$ range for SAX
J1711.6$-$3808, and a maximal spin value is only excluded at the
$2\sigma$ level of confidence.  Similarly, fits to XTE J1908$+$094
only exclude minimal and maximal spin at the $2\sigma$ level of
confidence.

The most important source of systematic error in our results may
derive from theoretical uncertainties in the ``effective'' ISCO.  We
have assumed that the inner radius of the disk is the ISCO as defined
by test particle orbits.  In an actual fluid disk, the effective inner
disk radius may differ slightly from the nominal ISCO.  The most
recent and sophisticated simulations that address this issue are
summarized in Reynolds \& Fabian (2008); that work is especially
relevant because the disk is taken to be {\it thin}.  (Prior
simulations considered hot, thick disks, which are unlikely to apply
in the regimes sampled by our data; see, e.g., Krolik, Hawley, \&
Hirose 2005.)  The surface density and ionization parameter of
orbiting gas is found to change sharply at the ISCO.  Based on the
results summarized in Reynolds \& Fabian (2008) and related work by
Shafee et al.\ (2008), a conservative estimate of the uncertainty in
the ISCO based on this and other simulations is $0.5~r_{g}$ for low
spin parameters (an uncertainty of about 20\% for very low spin; see
Miniutti \& Fabian 2009); the uncertainty is less for high spin
parameters.

Prior work on GX~339$-$4 is helpful in estimating systematic errors
related to the state in which a source is observed.  Miller et al.\
(2008) find that the spin parameter derived from fits to a single
spectrum from a given state differ from those derived by jointly
multiple spectra from three states by only 4\%.  Prior work on this
source is also helpful in evaluating systematic effects due to
different disk reflection models.  Reis et al.\ (2008) fit spectra of
GX 339$-$4 with a new disk reflection model, in which the atmosphere
is directly influenced by blackbody emission in the midplane.  Despite
the different reflection models employed by Miller et al.\ (2008) and
Reis et al.\ (2008), the spin parameters derived are fully consistent.
At least in the case of stellar-mass black holes with high inferred
spin parameters, then, modest differences in disk reflection models
do not strongly affect spin results.

Detector calibration is another potential source of systematic
error.  In the case of disk reflection features, however, the degree
of systematic error is likely to be small.  Silicon-based detectors
often suffer a sharp change in effective area around 2~keV, but these
features can be modeled and should not strongly affect the 4--8~keV
range.  Calibration uncertainties also typically arise below 0.7 keV
in CCD spectra; here again, uncertainties in this region will not
affect fits at much higher energy.  The response of xenon-based gas
detectors may be a larger source of systematic error: the Xe L3 edge
at 4.78~keV -- if not modeled correctly -- can affect fits to
relativistic iron lines. 

Most of the strong constraints that we have obtained come from CCD
spectra; the results obtained from XTE J1550$-$564 are the exception.
This suggests that the resolution afforded by CCD spectrometers plays
an important role in obtaining black hole spin constraints.  The
energy resolution of most gas spectometers is $E/dE \sim 6$.  At
6~keV, that is a resolution of 1~keV, or about 0.17$c$ -- a large
fraction of the velocity of matter orbiting at the ISCO.  Only
extremely strong emission lines will permit strong constraints in data
obtained with gas spectrometers.

\subsection{On the Role of the Disk Continuum}
Fits obtained with the simple ``diskbb'' disk continuum model (Mitsuda
et al.\ 2004) are, in general, statistically comparable to fits made
using the ``kerrbb'' model (Li et al.\ 2005) and yield similar spin
parameters, though in this case the spin is only measured through the
relativistic reflection signatures (see Table 2 and Table 3).  A
notable exception may be 4U 1543$-$475: stronger spin constraints are
derived when the ``kerrbb'' model is used.  We have assumed that the
inclination of the inner disk in 4U 1543$-$475 equals the binary
inclination, which is tightly constrained (Park et al.\ 2004).  The
fact that ``kerrbb'' includes the inclination as a free parameter --
which is coupled to the inclination in the blurred reflection model --
may account for the improvement.

The dominant role of the disk reflection spectrum is consistent with
some simple expectations.  Whereas a disk line bears the imprints of
gravitational red-shifts and Doppler shifts, disk continua do not
manifest equally unambiguous signatures of black hole spin.  An
especially high temperature -- perhaps indicative of a high efficiency
and black hole spin -- can instead be explained by a high mass
accretion rate.  Similarly, a small emitting area can potentially be
explained in terms of incorrectly accounting for scattering of the
thermal continuum in the disk.

Systematic uncertainties in necessary physical inputs may also have
kept the disk continuum from driving most of the spin values.  Most of
the uncertainties in black hole mass and distance listed in Table 1
are rather large, in a fractional sense.  Moreover, different cameras
aboard the same observatory do not necessarily measure the same flux
level.  In our fits to the two {\it ASCA}/GIS spectra of XTE
J1550$-$564, for instance, we have handled such systematics by
allowing the cameras to find different thermal disk properties apart
from the black hole spin parameter (values generally differ by less
than 20\%).  The procedure we have adopted is a compromise between the
principle of using physical models and the need to
acknowledge real observational limitations.

Although the simpler disk continuum model generally produced
equivalent results, we regard the results obtained using the more
physical continuum models as more definitive for three reasons.
First, the physical models do not give fits that are dramatically
inferior, in a statistical sense.  Second, the physical models do not
appear to skew the spin parameters in a particular direction.
Finally, and perhaps most importantly, the physical disk models do
attempt to account for important processes and spin explicitly, and so
enable a degree of self-consistency in the total spectral model.

\section{Discussion}
We have analyzed X-ray spectra from eight stellar-mass black holes.
The spectra were modeled with a combination of a
relativistically-blurred disk reflection spectrum and a new disk
continuum model in which spin is a variable parameter.  Spin estimates
were jointly derived by linking the spin parameters in the reflection
and continuum models.  This is the first effort to estimate spin
parameters in a number of sources by combining these independent
spectral diagnostics.  Our results suggest that stellar-mass black
holes may have a range of spin parameters.  Below, we discuss the
implications of this finding for black hole creation events,
relativistic jet production, and efforts to probe the innermost
relativistic regime around black holes.

A black hole needs to accrete a large fraction of its mass to reach
maximal spin (Volonteri et al.\ 2005).  In the case of stellar-mass
black holes, then, spin parameters should be determined primarily by
the supernova or gamma-ray burst event that creates the black hole.
Theoretical investigations into nascent black hole spin parameters,
set by a single collapse event, have focused on supermassive stars in
the early universe.  Only considering the collapse event, a spin of $a
\simeq 0.75$ is expected (Shibata \& Shapiro 2002); with additional
considerations, a spin as high as $a \leq 0.93$ may be possible
(Gammie et al.\ 2004).  Our results are nominally at odds with these
predictions.  However, as noted by Heger \& Woosley (2002) and Gammie
et al.\ (2004), the mass, angular momentum, metallicity, and magnetic
field structure in progenitor stars are all important in determining
its final spin, and considerable theoretical uncertainties remain in
modeling this problem.  More theoretical and observational work is
needed before it is clear that collapse/explosion models are incorrect
or that some black holes were formed in exotic circumstances.

The most extreme result we have obtained may be the near-zero spin for
Cygnus X-1.  The {\it XMM-Newton} spectra that we analyzed were
obtained in a ``high'' state (called the ``very high'' state in other
black holes); this makes it very unlikely that the disk was far from
the ISCO when the source was observed.  Additional observational work
with {\it Suzaku} is needed to confirm a low spin parameter.  The fact
that some SNe leave behind neutron stars with modest magnetic fields,
while others give rise to magnetars, is one important indication that
stellar explosions can leave behind objects with very different
properties.  Cygnus X-1 is the only high-mass binary in our sample,
and this property may be important in determining its spin.  Indeed,
Cygnus X-1 may be even more special: Mirabel \& Rodrigues (2003)
suggest that Cygnus X-1 may have formed in an unusual supernova with
very little mass loss, owing to the absence of a supernova remnant and
its low space velocity.

A connection between spin and jets is anticipated theoretically
(e.g. Blandford \& Znajek 1977), and our results would appear to hint
at a connection: GX 339$-$4, GRO~J1655$-$40, and XTE J1550$-$564 are
all relativistic jet sources where velocities above $0.9c$ have been
inferred (Gallo et al.\ 2004, Hjellming \& Rupen 1995, Hannikainen et
al.\ 2001), and all of them are found to have relatively high spin
values.  A high spin parameter is also implied in XTE J1650$-$500, but
no relativistic jet was detected in this source.  The absence of such
a detection may be due to the limited angular resolution of
instruments in the Southern Hemisphere.  Cygnus X-1 may again be an
important exception: our fits imply a very low spin parameter, yet
Cygnus X-1 powers a jet with $v/c \geq 0.6$ (Stirling et al.\ 2001).

Taken as a whole, our results only weakly support a link between
spin and jet power.  Future observations of black hole transients in
both X-ray and radio bands may be able to build a sample from which
stronger conclusions can be drawn.  Not every black hole in our sample
was observed intensively in the radio band or at high angular
resolution; relativistic jets could have been missed in some sources.
Moreover, spin may not be the only parameter important in producing
relativistic jets, especially since jets are not seen in all black
hole states.  Garcia et al.\ (2003) found that relativistic jet
sources tend to be those with long orbital periods.  It might be the
case that a parameter such as the absolute mass accretion rate through
the disk is important.

It is worth noting that in the cases where high spin parameters are
required by our spectral models, the line emissivity index is found to
be very steep ($q\sim5$).  Both a high spin parameter and a steep
emissivity index would serve to concentrate the reflected emission in
the very innermost part of the disk.  High spin parameters are
therefore required despite steep emissivity profiles, not because of
steep emissivity profiles.  Physically, an emissivity of $q>3$ may
imply that the source of hard X-ray flux in these systems may be very
compact, and may radiate anisotropically.  Prior work has suggested
that the compact coronae implied may be consistent with the base of a
jet (e.g. Miller et al.\ 2004b).  This inner accretion flow geometry
is similar to that described in models for line and continuum
variability that invoke gravitational light bending close to a
spinning black hole (see Miniutti \& Fabian 2004).  Gravitational
light-bending may be partially responsible for concentrating emission
centrally.  Investigations of line variability in Seyfert AGN with the
{\it International X-ray Observatory} (IXO) will be able to provide much
stronger evidence of gravitational light bending.

The new results reported in this work are broadly consistent with prior
estimates of black hole spin using line models with fixed spin
parameters.  For instance, in the case of XTE J1650$-$500, prior fits
to spectra obtained with {\it XMM-Newton} and {\it BeppoSAX} indicated
inner disk radii of $2 r_{g}$ or less (Miller et al.\ 2002; Miniutti,
Fabian, \& Miller 2004).  Prior fits to {\it Chandra} and {\it
XMM-Newton} spectra of GX~339$-$4 strongly suggested a high spin
parameter (Miller et al.\ 2004a, 2004b, 2006); later fits to {\it
XMM-Newton} and {\it Suzaku} spectra with variable-spin disk
reflection models yielded results consistent with those reported here
(Miller et al.\ 2008; see also Reis et al.\ 2008).

The spin parameters that we have obtained for 4U 1543$-$475 and
GRO~J1655$-$40 do not fully agree with recent modeling of the
accretion disk continuum alone (Shafee et al.\ 2006; also see Zhang,
Cui, \& Chen 1997).  Whereas we measure $a = 0.3(1)$ for 4U
1543$-$475, Shafee et al.\ (2006) find $a = 0.75-0.85$.  And whereas
we measure $a = 0.98(1)$ for GRO J1655$-$40, Shafee et al.\ (2006)
give $a = 0.65-0.75$.  The differences may be partially derive from
lingering systematic errors and differing analysis procedures.  As
noted previously, spin estimates based on the continuum require an
absolute flux measurement, and so require accurate knowledge of the
mass and distance to the source, and incur systematics due to
uncertainties in the absolute flux calibration of a given instrument.
In a number of fits, Shafee et al.\ (2006) only consider {\it RXTE}
spectra below 8~keV, and in other cases the line and reflection
continuum is modeled with a Gaussian and (unphysical) smeared edge.
The disk reflection spectrum, which is less subject to systematic
uncertainties, appears to drive the spin constraints we have obtained.
In the broadest sense, future work aimed at resolving the differences
such as $a=0.3(1)$ versus $a=$0.75-0.85 (for 4U 1543$-$475) and
$a=0.98(1)$ versus $a=$0.65-0.75 (for GRO~J1655$-$40) is a welcome
prospect and marks a turning point for studies of stellar-mass black
holes.

The results we have obtained underscore the urgent need to obtain
X-ray spectra of multiple sources in multiple states, with moderate or
high resolution spectrometers.  Observing multiple states and jointly
fitting resultant spectra to require a common spin parameter (Miller
et al.\ 2008, Reis et al.\ 2008) ensures that any variations in the
ionization and/or structure of the accretion disk are treated
explicitly.  {\it Chandra} and {\it XMM-Newton} are both well-suited
to this aim; the broad-band spectroscopy and high throughput, fast CCD
read-out of {\it Suzaku} mean it is an exceptional tool for disk
reflection studies.  Future missions, including {\it Astro-H} (NeXT)
and the {\it IXO} will be able to measure spin in stellar-mass black
holes with unprecedented precision, if high source fluxes can be
accommodated.  The knowledge gained in understanding the spin
distribution of stellar-mass black holes can be transferred to studies
of supermassive black hole spins.

Our results also highlight the need to measure more black hole masses
through dynamics, and to push current and future mass constraints to
higher precision.  Even in this initial attempt at a systematic
analysis, it is already clear that uncertainties in binary parameters
inhibit stronger conclusions.  If we are to understand how supernovae
and GRBs might produce black holes with differing spin parameters,
robust measurements of the remnant black hole mass, its companion
mass, and other system parameters will be important factors.
Especially when a black hole lies in a region with high line-of-sight
column density, rapid IR observations may help to identify
counterparts and enable later dynamical studies.  The properties of
GRS~1915$+$105 have not yet been measured precisely (Greiner, Cuby, \&
McCaughrean et al.\ 2001), but the fact that measurements are possible
given a acolumn density of $N_{H} \simeq 4\times 10^{22}~ {\rm atoms}~
{\rm cm}^{-2}$ (Dickey \& Lockman 1990) signals that there is a way
forward in such cases.

\section{Conclusions}

We have fitted relativistic disk reflection and disk continuum models
to a number of stellar-mass black holes.  A broad range of spin
parameters is measured.  This implies a fundamental diversity in the
GRB/SNe events that are thought to create stellar-mass black holes.
The black holes with the highest spin parameters are those wherein the
most relativistic jets have been observed in radio bands, providing
modest support for a connection between spin and jets.  With a larger
sample, the influence of binary system parameters (if any) on black
hole spin may become apparent.

\vspace{0.1in}
We acknowledge the anonymous referee for comments that improved this
manuscript.We wish to thank Jean in 't Zand for generously making the
{\it BeppoSAX} spectra of SAX J1711.6$-$3808 and XTE J1908$+$094
available.  We acknowledge helpful conversations with Charles Gammie,
Julian Krolik, and Cole Miller.  J. M. M. thanks Keith Arnaud, Brian
Irby, and Roy Bonser for help with XSPEC and HEASOFT.
C. S. R. acknowledges support from the NSF under grant AST06-07428.
G. M. thanks the Spanish Ministerio de Ciencia e Innovaci\'on and the
CSIC for support through a Ram\'on y Cajal contract.  This work made
use of the tools and information available through the HEASARC
facility, operated for NASA by GSFC.

\clearpage

\centerline{~\psfig{file=f1.ps,width=3.2in,angle=-90}~}
\figcaption[h]{\footnotesize The plot above shows the {\it RXTE}
spectrum of 4U 1543$-$475 fit with a phenomenological disk plus
power-law model.  The 4--7 keV range was ignored when fitting the
spectrum to best illustrate the relativistic iron line.}
\medskip

\centerline{~\psfig{file=f2.ps,width=3.2in,angle=-90}~}
\figcaption[h]{\footnotesize The plot above shows the {\it RXTE}
spectrum of 4U 1543$-$475 fit with a relativistically-blurred
reflection spectrum and the ``kerrbb'' disk continuum model.  The spin
parameters in the disk reflection and continuum spectra were linked.}
\medskip

\clearpage

\centerline{~\psfig{file=f3.ps,width=3.2in,angle=-90}~}
\figcaption[h]{\footnotesize The plot above shows the {\it ASCA}
spectra of XTE J1550$-$564 fit with a phenomenological disk plus
power-law model.  The 4--7 keV range was ignored when fitting the
spectra to best illustrate the relativistic iron lines.}
\medskip

\centerline{~\psfig{file=f4.ps,width=3.2in,angle=-90}~}
\figcaption[h]{\footnotesize The plot above shows the {\it ASCA}
spectra of XTE J1550$-$564 fit with a relativistically-blurred
reflection spectrum and the ``kerrbb'' disk continuum model.  The spin
parameters in the disk reflection and continuum spectra were linked.}
\medskip

\clearpage

\centerline{~\psfig{file=f5.ps,width=3.2in,angle=-90}~}
\figcaption[h]{\footnotesize The plot above shows the {\it XMM-Newton}
spectrum of XTE J1650$-$500 fit with a phenomenological disk plus
power-law model.  The 4--7 keV range was ignored when fitting the
spectrum to best illustrate the relativistic iron line.  Features
around 2 keV are instrumental.}
\medskip

\centerline{~\psfig{file=f6.ps,width=3.2in,angle=-90}~}
\figcaption[h]{\footnotesize The plot above shows the {\it XMM-Newton}
spectrum of XTE J1650$-$500 fit with a relativistically-blurred
reflection spectrum and the ``kerrbb'' disk continuum model.  The spin
parameters in the disk reflection and continuum spectra were linked.}
\medskip

\clearpage

\centerline{~\psfig{file=f7.ps,width=3.2in,angle=-90}~}
\figcaption[h]{\footnotesize The plot above shows the {\it RXTE}
spectra of GRO~J1655$-$40 fit with a phenomenological disk plus
power-law model.  The 4--7 keV range was ignored when fitting the
spectrum to best illustrate the relativistic iron line.}
\medskip

\centerline{~\psfig{file=f8.ps,width=3.2in,angle=-90}~}
\figcaption[h]{\footnotesize The plot above shows the {\it RXTE}
spectra of GRO J1655$-$40 fit with a relativistically-blurred
reflection spectrum and the ``kerrbb'' disk continuum model.  The spin
parameters in the disk reflection and continuum spectra were linked.}
\medskip

\clearpage

\centerline{~\psfig{file=f9.ps,width=3.2in,angle=-90}~}
\figcaption[h]{\footnotesize The plot above shows the {\it XMM-Newton}
spectrum of GX 339$-$4 fit with a phenomenological disk plus
power-law model.  The 4--7 keV range was ignored when fitting the
spectrum to best illustrate the relativistic iron line.  Features
around 2 keV are instrumental.}
\medskip

\centerline{~\psfig{file=f10.ps,width=3.2in,angle=-90}~}
\figcaption[h]{\footnotesize The plot above shows the {\it XMM-Newton}
spectrum of GX~339$-$4 fit with a relativistically-blurred
reflection spectrum and the ``kerrbb'' disk continuum model.  The spin
parameters in the disk reflection and continuum spectra were linked.}
\medskip

\clearpage

\centerline{~\psfig{file=f11.ps,width=3.2in,angle=-90}~}
\figcaption[h]{\footnotesize The plot above shows the {\it BeppoSAX}
spectrum of SAX J1711.6$-$3808 fit with a phenomenological power-law
model.  The 4--7 keV range was ignored when fitting the spectrum to
best illustrate the relativistic iron line.}
\medskip

\centerline{~\psfig{file=f12.ps,width=3.2in,angle=-90}~}
\figcaption[h]{\footnotesize The plot above shows the {\it BeppoSAX}
spectrum of SAX J1711.6$-$3808 fit with a relativistically-blurred
reflection model.  }
\medskip

\clearpage

\centerline{~\psfig{file=f13.ps,width=3.2in,angle=-90}~}
\figcaption[h]{\footnotesize The plot above shows the {\it BeppoSAX}
spectra of XTE J1908$+$094 fit with a phenomenological power-law
model.  The 4--7 keV range was ignored when fitting the spectrum to
best illustrate the relativistic iron line.}
\medskip

\centerline{~\psfig{file=f14.ps,width=3.2in,angle=-90}~}
\figcaption[h]{\footnotesize The plot above shows the {\it BeppoSAX}
spectra of XTE J1908$+$094 fit with a relativistically-blurred
reflection model.  }
\medskip

\clearpage

\centerline{~\psfig{file=f15.ps,width=3.2in,angle=-90}~}
\figcaption[h]{\footnotesize The plot above shows the {\it XMM-Newton}
spectra of Cygnus X-1 fit with phenomenological disk plus
power-law models.  The 4--7 keV range was ignored when fitting the
spectra to best illustrate the relativistic iron lines.  Features
around 2 keV are instrumental.}
\medskip

\centerline{~\psfig{file=f16.ps,width=3.2in,angle=-90}~}
\figcaption[h]{\footnotesize The plot above shows the {\it XMM-Newton}
spectra of Cygnus X-1 fit with a relativistically-blurred
reflection spectrum and the ``kerrbb'' disk continuum model.  The spin
parameters in the disk reflection and continuum spectra were linked.}
\medskip

\clearpage

\centerline{~\psfig{file=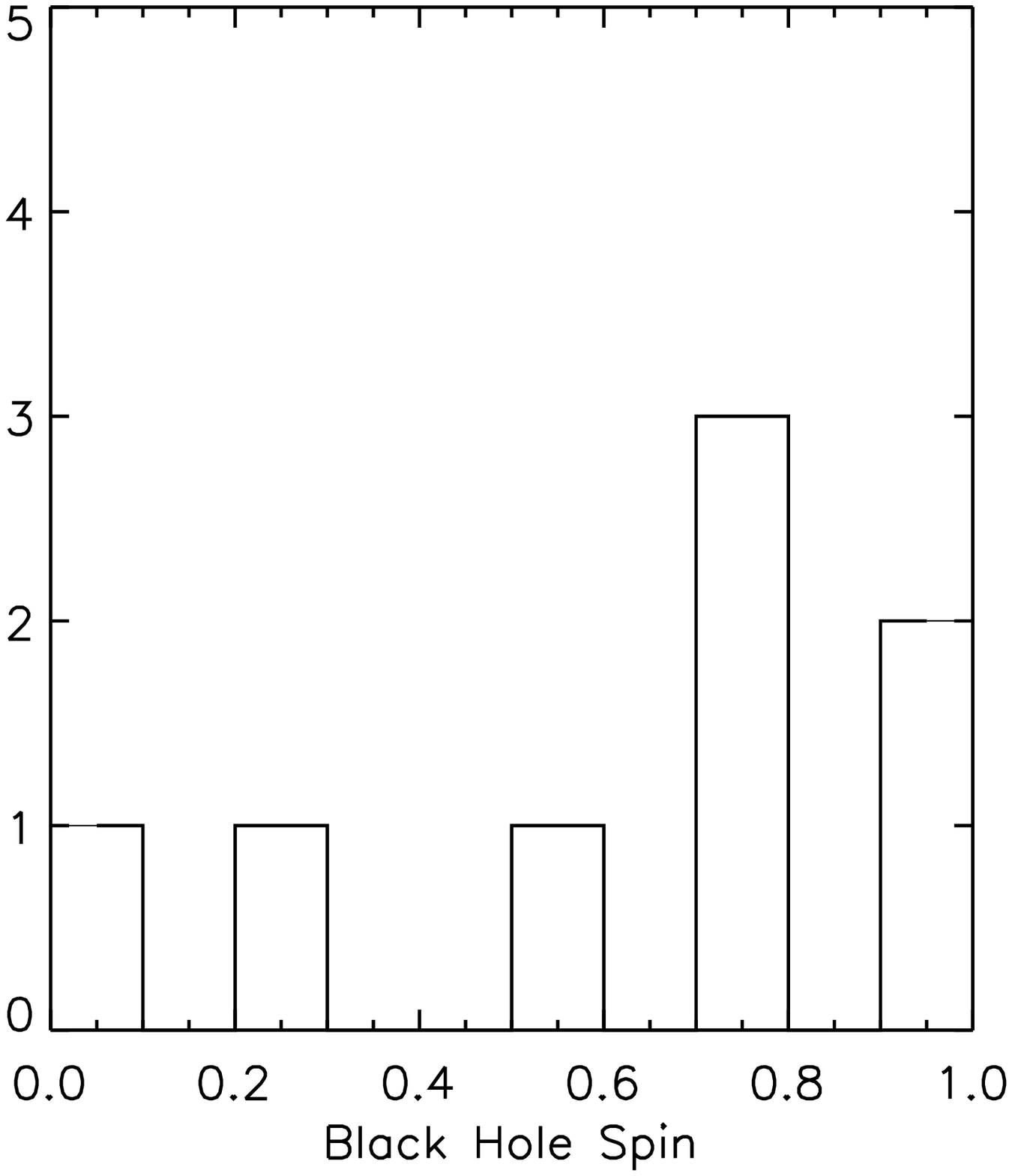,width=5.0in}~}
\figcaption[h]{\footnotesize The histogram above plots the
distribution of black hole spin parameters obtained through our fits.
Please see the text for details and caveats.}
\medskip

\clearpage

\centerline{~\psfig{file=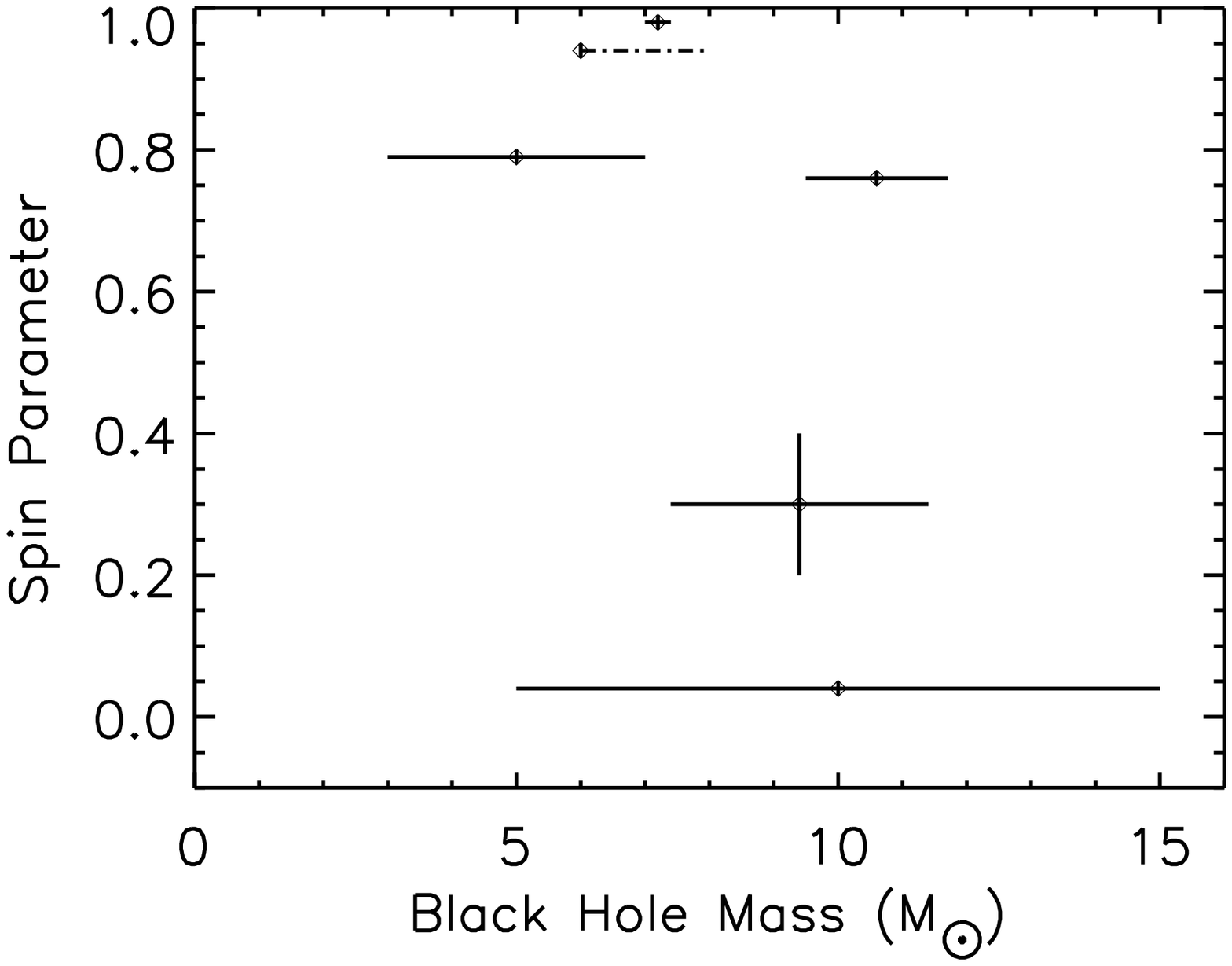,width=5.0in}~}
\figcaption[h]{\footnotesize The plot above shows our derived
$1\sigma$ spin constraints versus black hole mass, for sources with
known masses.  The dashed, single-sided error above denotes that only
a lower mass limit has been obtained for GX~339$-$4.  Please see
Section 2 for more details.}
\medskip

\clearpage

\centerline{~\psfig{file=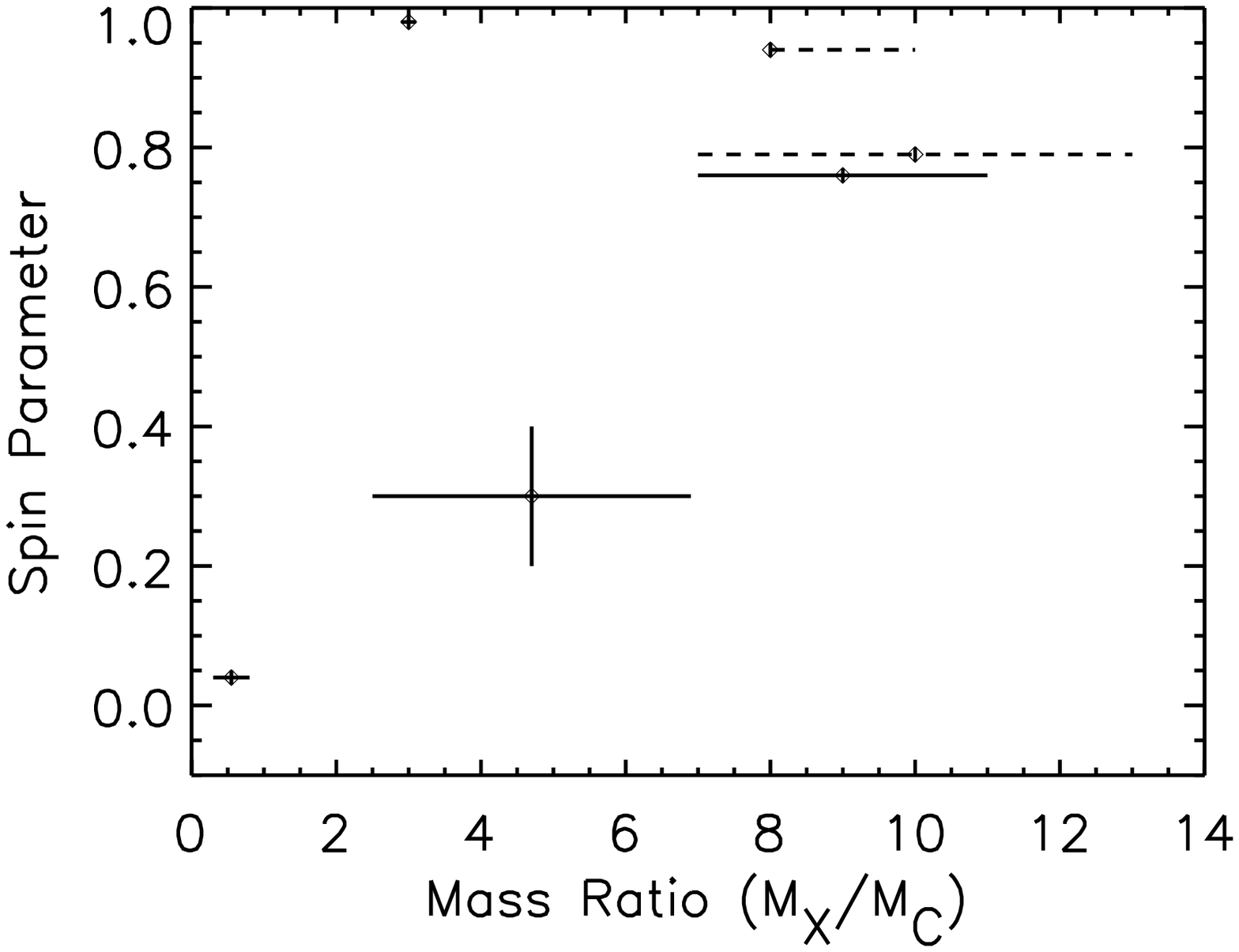,width=5.0in}~}
\figcaption[h]{\footnotesize The plot above shows our derived
$1\sigma$ spin constraints versus the ratio of the black hole mass to
the companion mass, for sources with known masses.  The dashed,
single-sided error above denotes that only a lower limit has been
obtained for the mass ratio in GX~339$-$4.  The dashed, double-sided
error denotes that the mass ratio in XTE J1650$-$500 is largely
unconstrained.}
\medskip

\clearpage

\centerline{~\psfig{file=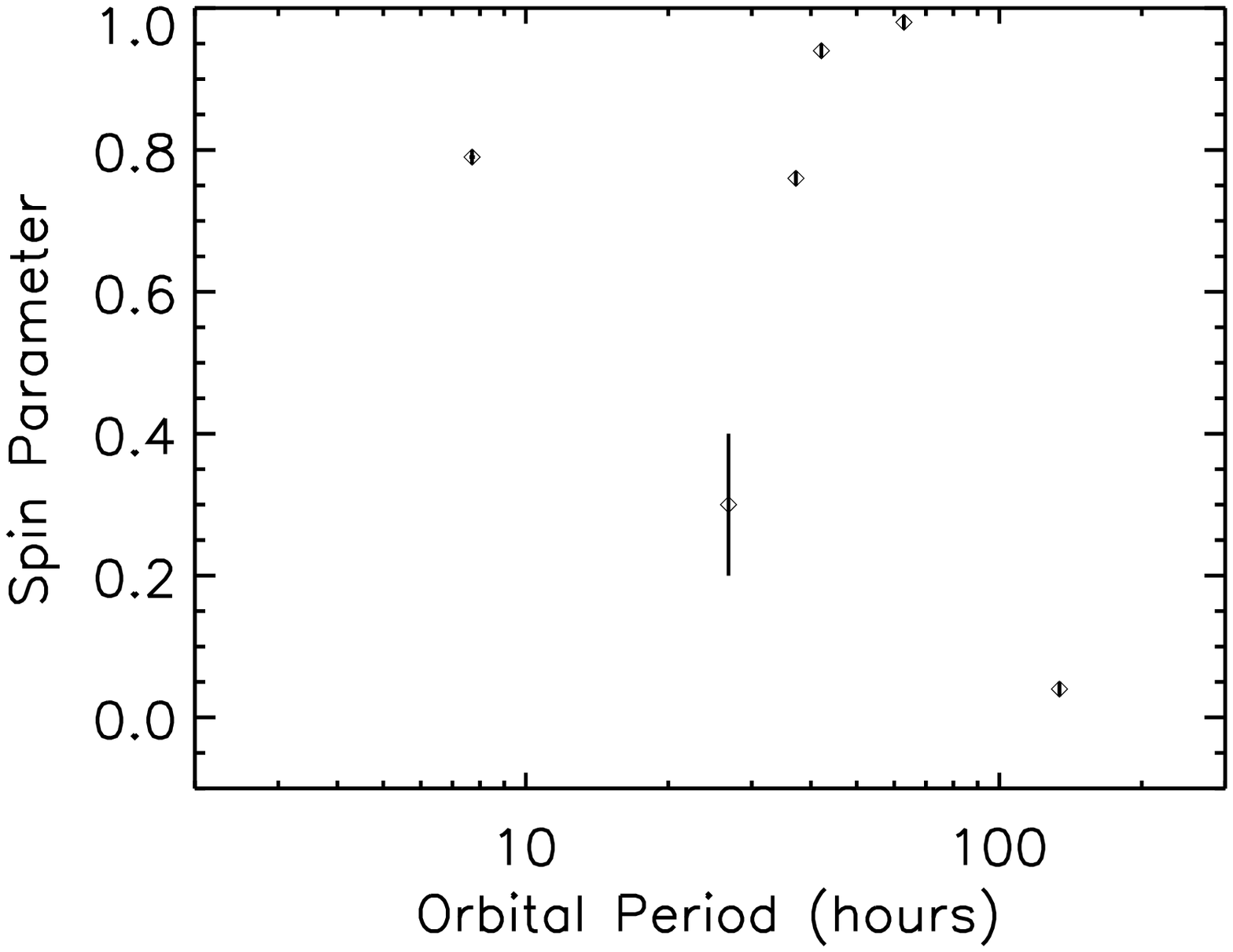,width=5.0in}~}
\figcaption[h]{\footnotesize The plot above shows our derived
$1\sigma$ spin constraints versus the orbital period of the binary,
for sources with known masses.  The errors on orbital period are
plotted but they are very small.}
\medskip

\clearpage

\centerline{~\psfig{file=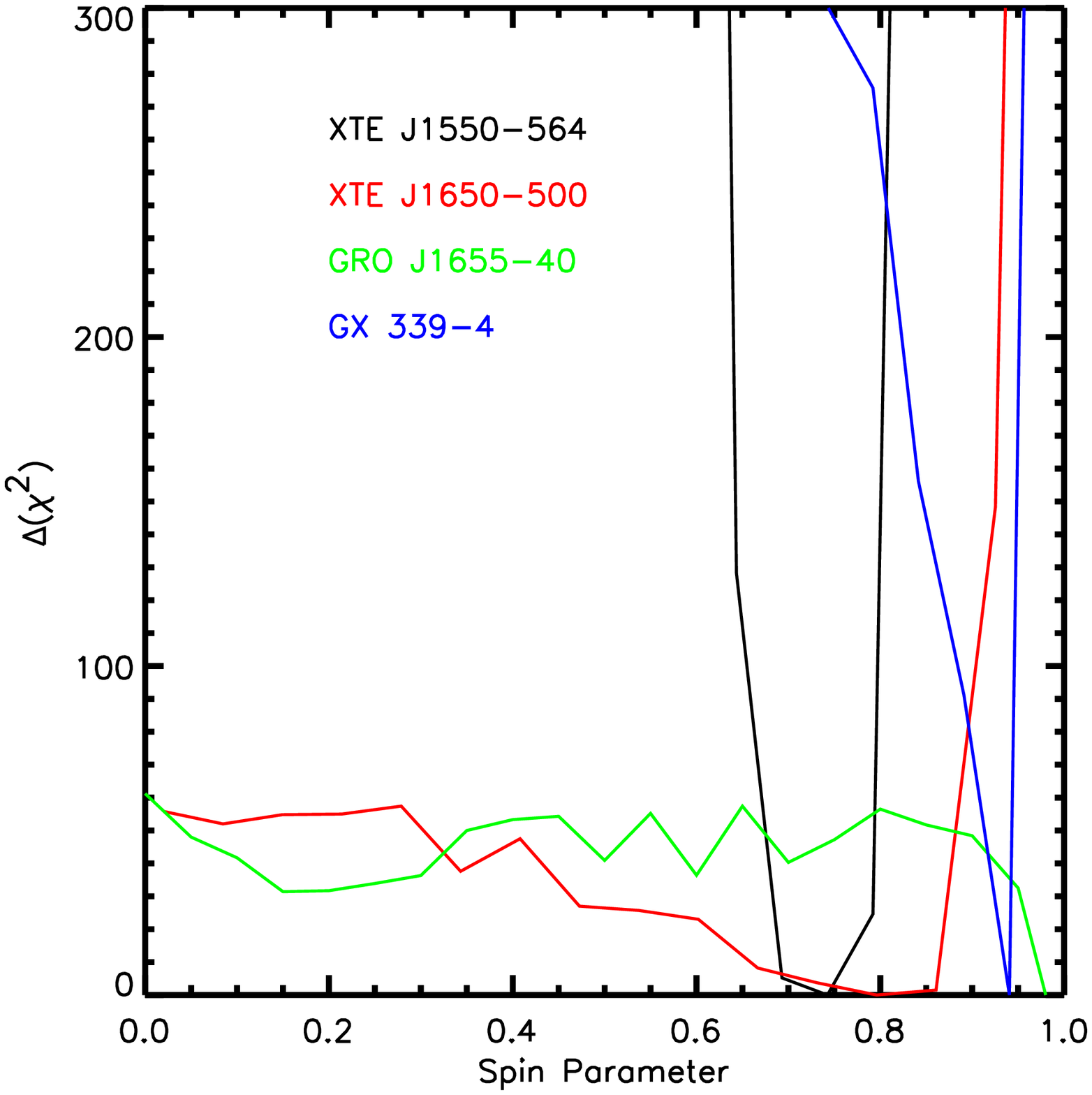,width=5.0in}~}
\figcaption[h]{\footnotesize The plot above shows the change in the
goodness-of-fit statistic as a function of the black hole spin
parameter, $a$.  Using the XSPEC ``steppar'' command, 20 evenly-spaced
value of $a$ were frozen and all other parameters were allowed to
float freely to find the best fit at that spin parameter.  For
clarity, only sources measured to have high spin at high confidence
are shown above.}
\medskip

\clearpage

\centerline{~\psfig{file=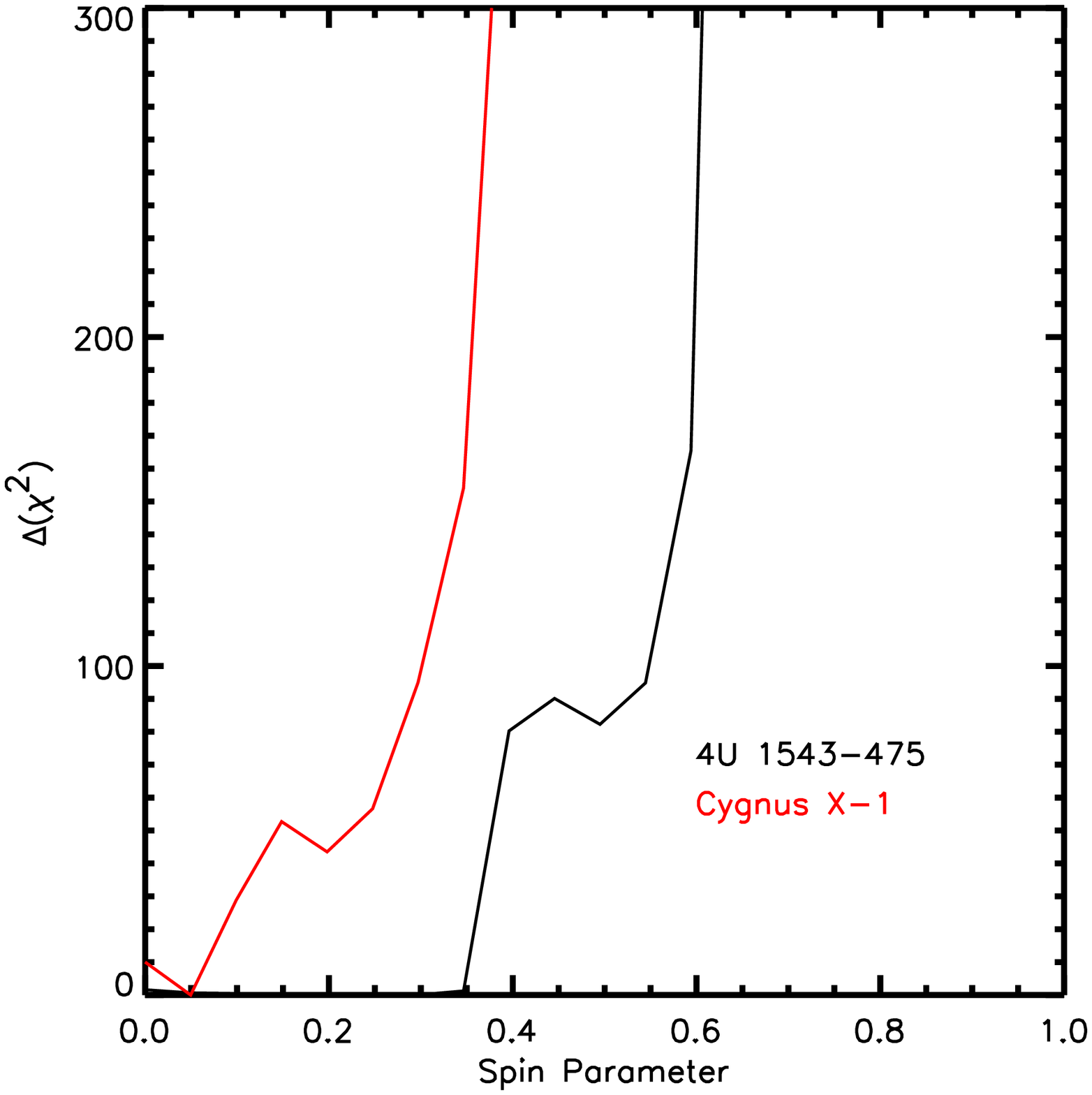,width=5.0in}~}
\figcaption[h]{\footnotesize The plot above shows the change in the
goodness-of-fit statistic as a function of the black hole spin
parameter, $a$.  Using the XSPEC ``steppar'' command, 20 evenly-spaced
value of $a$ were frozen and all other parameters were allowed to
float freely to find the best fit at that spin parameter.  For
clarity, only sources measured to low high spin at high confidence
are shown above.}
\medskip

\clearpage

\centerline{~\psfig{file=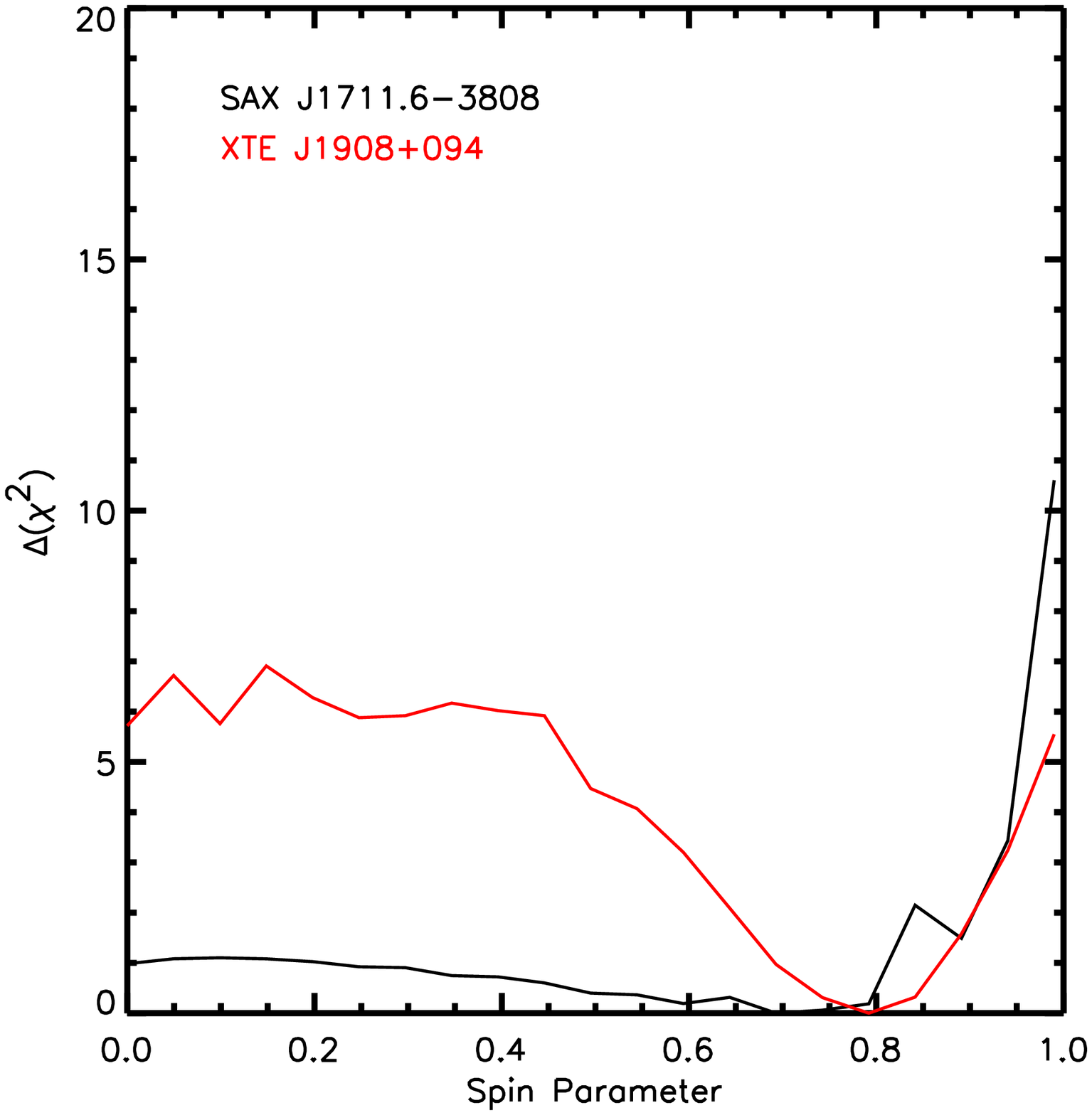,width=5.0in}~}
\figcaption[h]{\footnotesize The plot above shows the change in the
goodness-of-fit statistic as a function of the black hole spin
parameter, $a$.  Using the XSPEC ``steppar'' command, 20 evenly-spaced
value of $a$ were frozen and all other parameters were allowed to
float freely to find the best fit at that spin parameter.  For
clarity, only sources for which spin is not confidently determined are
shown above.}

%----------------------------------------------------------------------

\clearpage

\begin{table}[htb!]
\caption{Black Hole Binary Parameters}
\begin{footnotesize}
\begin{center}
\begin{tabular}{lllllll}
Source & ${\rm M}_{\rm BH}$ & ${\rm M}_{\rm C}$ & $q$ & $\theta$ & ${\rm P}_{\rm orb}$ & Dist.\\
~ & ${\rm M}_{\odot}$ & ${\rm M}_{\odot}$ & ~ & (deg.) & (hours) & (kpc) \\
\tableline
4U 1543$-$475$^{a}$ & $9(2)$ & 3(1) & 5(2) & 21(1) & 26.8(1) & 8(1) \\
XTE J1550$-$564$^{b}$ & 11(1) & 1.3(1) & 9(2) & 50-80 & 37.2(2) & 3--8 \\
XTE J1650$-$500$^{c}$ & 5(2) & 0.3--0.7 & $\sim$10 & $\sim$50 & 7.6(2) & 5--11 \\
GRO J1655$-$40$^{d}$ & 7.0(2) & 2.3(1) & 3.0(2) & 77(8) & 62.9(1) & 3.2(9) \\
GX 339$-$4$^{e}$ & $\geq6$ & $\geq$0.8 & $\geq$8 & -- & 42.1(1) & 8--15 \\
Cygnus X-1$^{f}$ & 10(5) & $\sim$18 & 0.6(3) & $\sim 35$ & 134.0(1) & 2.5(5) \\
\tableline
\end{tabular}
%\vspace*{\baselineskip}~\\ 
\end{center} 
\tablecomments{Fundamental parameters for the black hole binaries
treated in this work are given above.  These parameters were used as
inputs to the ``kerrbb'' disk continuum model when fitting the X-ray
spectra in this paper.  Parameters listed with a wave-like mark are
very uncertain and were not used to bound parameters in fits with the
``kerrbb'' model.  Instead, those parameters floated freely in all
fits.  Please see Section 2 for additional details.
$^{a}$ Parameters taken from Park et al.\ (2004).
$^{b}$ Parameters taken from Orosz et al.\ (2002).
$^{c}$ Parameters taken from Orosz et al.\ (2004).
$^{d}$ Parameters taken from Orosz \& Bailyn (1997) and Hjellming \& Rupen (1995).
$^{e}$ Parameters taken from Hynes et al.\ (2003, 2004) and Munoz-Darias, Casares, \& Martinez-Pais (2008).
$^{f}$ Parameters taken from Herrero et al.\ (1995).
}
\vspace{-1.0\baselineskip}
\end{footnotesize}
\end{table}

\clearpage

\begin{table}[htb!]
\caption{Fully Relativistic Spectral Fits}
\begin{footnotesize}
\begin{center}
\begin{tabular}{lllllllllllll}
Source & $a$ & $i$ & $q$ & $N_{H}$ & $\Gamma$ & $R$ & log($\xi$) & $N_{\rm refl.}$ & $N_{\rm disk}$ & $\dot{M}_{disk}$ & $\nu$ & $\chi^{2}/\nu$ \\
~ & ~ & (deg.) & ~ & $(10^{21}~{\rm cm}^{-2})$ & ~ & ~ & ~ & ($10^{-26}$) & ~ & ($10^{18}~{\rm g}~{\rm s}^{-1}$)x & ~ & ~ \\

\tableline

4U 1543$-$475 & $0.3(1)$ & $22_{-1}$ & $3.0^{+0.1}$ & 4.0 & $2.48(5)$ & $2.0_{-0.1}$ & 4.5(1) & 3.88(5) & 0.72(2) & 1.62(2) & 55 & 1.15 \\

XTE J1550$-$564 & 0.76(1) & $50^{+1}$ & $5.0_{-0.1}$ & 6.67(1) & 1.79(1) & $2.0_{-0.1}$ & 3.86(2) & 1.41(2) & 0.23(1) & 0.70(1) & 1493 & 1.23 \\

XTE J1650$-$500 & 0.79(1) & 45(1) & $4.9(1)$ & 5.58(1) & 1.97(1) & 0.56(1) & 3.80(4) & 4.96(1) & 49.5(5) & $0.007^{+0.005}_{-0.002}$ & 1853 & 1.10 \\

GRO J1655$-$40 & 0.98(1) & $69^{+1}$ & $5.0_{-0.1}$ & 20.5(5) & 2.60(2) & 0.18(2) & $4.3^{+0.2}_{-0.5}$ & 140(9) & 0.25(4) & 0.18(4) & 61 & 0.91 \\

GX 339$-$4 & 0.94(2) & 29(2) & 4.9(1) & 5.7(1) & 2.74(5) & $2.0_{-0.2}$ & 4.8(2) & 1.6(2) & 0.67(5) & 1.09(3) & 1816 & 1.54 \\

SAX J1711.6$-$3808 & $0.6^{+0.2}_{-0.4}$ & 43(5) & $3.0^{+0.1}$ & 25.8(7) & 1.69(2) & 0.25(1) & 3.0(1) & 6.5(2) & -- & -- & 40 & 0.93 \\

XTE J1908$+$094 & $0.75(9)$ & 45(8) & $3.0^{+0.1}$ & 21.5(5) & 1.89(7) & 0.17(1) & 3.0 & 9(1) & -- & -- & 297 & 1.51 \\

Cygnus X-1 & $0.05(1)$ & 30(1) & $3.0^{+0.1}$ & 6.84(2) & 2.74(1) & 1.3(1) & 3.5(1) & 290(5) & 0.31(2) & 0.96(1) & 3687 & 1.78 \\

\tableline
\end{tabular}
%\vspace*{\baselineskip}~\\ 
\end{center} 
\tablecomments{The parameters listed are derived from fits to our
stellar-mass black hole spectra with a blurred CDID reflection model
and the ``kerrbb'' continuum model.  Spin values and inclinations in
the blurred reflection and disk continuum models were linked.  To
employ the ``kerrbb'' model, a number of parameters must be fixed or
constrained a priori (e.g. distance, black hole mass, disk atmosphere
hardening factor, etc.); for each source, please see the text for
related details.  The mass accretion rate given avove is an
``effective'' mass accretion rate; see Li et al.\ (2005) for details.
The errors listed above are 1$\sigma$ statistical errors, derived
using the ``error'' command in XSPEC.  Symmetric errors are given in
parentheses; where one digit appears in parentheses it is the error in
the last digit of the parameter value.  SAX~J1711.6$-$3808 and
XTE J1908$+$094 did not require disk continuum components.  }
\vspace{-1.0\baselineskip}
\end{footnotesize}
\end{table}

\begin{table}[htb!]
\caption{Spectral Fits with a Simple Disk Continuum}
\begin{footnotesize}
\begin{center}
\begin{tabular}{lllllllllllll}
Source & $a$ & $i$ & $q$ & $N_{H}$ & $\Gamma$ & $R$ & log($\xi$) & $N_{\rm refl.}$ & $kT$ & $N_{\rm disk}$ & $\nu$ & $\chi^{2}/\nu$ \\
~ & ~ & (deg.) & ~ & $(10^{21}~{\rm cm}^{-2})$ & ~ & ~ & ~ & ($10^{-26}$) & (keV) & ($10^{3}$) & ~ &  ~ \\

\tableline

4U 1543$-$475 & $0.3^{+0.2}_{-0.3}$ & $22_{-1}$ & $3.1(1)$ & 4.0 & $2.50(5)$ & $2.0_{-0.1}$ & 4.5(1) & 0.41(4) & $0.55(1)$ & 9.40(2) & 55 & 1.17 \\

XTE J1550$-$564 & 0.78(2) & 50(1) & $5.0_{-0.1}$ & 6.70(1) & 1.92(1) & 1.1(1) & 3.92(6) & 2.08(1) & 0.66(1) & 1.04(1) & 1496 & 1.20 \\

XTE J1650$-$500 & 0.87(1) & 47(1) & $5.0_{-0.1}$ & 5.30(1) & 1.96(1) & 0.62(2) & 3.86(1) & 4.05(1) & 0.31(1) & 49.4(1) & 1855 & 1.10 \\

GRO J1655$-$40 & 0.94(3) & 70(1) & $4.1^{+0.3}_{-0.9}$ & 17.6(1) & 2.64(1) & 0.42(5) & 3.2(2) & 180(30) & 1.55(1) & 0.13(1) & 61 & 0.68 \\

SAX J1711.6$-$3808 & $0.6^{+0.2}_{-0.4}$ & 43(5) & $3.0^{+0.1}$ & 25.8(7) & 1.69(2) & 0.25(1) & 3.0(1) & 6.5(2) & -- & -- & 40 & 0.93 \\

XTE J1908$+$094 & $0.75(9)$ & 45(8) & $3.0^{+0.1}$ & 21.5(5) & 1.89(7) & 0.17(1) & 3.0 & 9(1) & -- & -- & 297 & 1.51 \\

Cygnus X-1 & $0.00^{+0.05}$ & 24(1) & $3.0^{+0.1}$ & 6.80(1) & 2.79(1) & 0.93(5) & 3.9(1) & 170(01) & 0.42(1) & 43.0(1) & 3689 & 2.01 \\

\tableline
\end{tabular}
\vspace*{\baselineskip}~\\ 
\end{center}
\tablecomments{ The parameters listed are derived from fits to our
stellar-mass black hole spectra with a blurred CDID reflection model
and the ``kerrbb'' continuum model.  The simple ``diskbb'' disk
continuum model was used; this model does not measure spin directly
and has only two variable parameters (temperature and flux).  Please
see the text for details on each source and spectrum.  Fits to GX
339$-$4 using this disk continuum are detailed in Miller et al.\
(2008).  The errors listed above are 1$\sigma$ statistical errors,
derived using the ``error'' command in XSPEC. Symmetric errors are
given in parentheses; where one digit appears in parentheses it is the
error in the last digit of the parameter value.  SAX~J1711.6$-$3808
and XTE J1908$+$094 did not require disk continuum components; values
from Table 2 are repeated here to enable comparisons.  }
\vspace{-1.0\baselineskip}
\end{footnotesize}
\end{table}


\medskip

\clearpage

\begin{references}

\reference{} Arnaud, K. A., and Dorman, B., 2000, XSPEC is available
via the HEASARC on-line service, provided by NASA/GSFC

\reference{} Ballantyne, D., Ross, R. R., \& Fabian, A. C., 2001,
MNRAS, 327, 10

\reference{} Balucinska-Church, M., \& Church, M. J., 2000, MNRAS, 311, 861

\reference{} Beckwith, K., \& Done, C., 2004, MNRAS, 352, 353

\reference{} Blandford, R. D., \& Znajek, R. L., 1977, MNRAS, 179, 433

\reference{} Blum, J. L., et al., 2009, ApJ, in preparation

\reference{} Brandt, W. N., et al., 1996, MNRAS, 283, 1071

\reference{} Brenneman, L. W., \& Reynolds, C. S., 2006, ApJ, 652, 1028

\reference{} Bhattacharyya, S., \& Strohmayer, T. E., 2007, ApJ, 664, L103

\reference{} Cackett, E., et al., 2008, ApJ, 674, 415

\reference{} Chartas, G., Kochanek, C. S., Dai, X., Poindexter, S., \& Garmire, G., 2008, ApJ, in press

\reference{} Cui, W., Ebisawa, K., Dotani, T., \& Kubota, A., 1998, ApJ, 493, L75

\reference{} Diaz Trgo, M., Parmar, A. N., Miller, J. M., Kuulkers,
E., Caballero-Garcia, M. D., et al., 2007, A\&A, 462, 657

\reference{} Dickey, J. M., \& Lockman, F. J., 1990, ARA\&A, 28, 215

\reference{} Dovciak, M., Karas, V., \& Yaqoob, T., 2004, ApJS, 153, 205

\reference{} Frank, J., King, A., \& Raine, D., 2002, in ``Accretion
Power in Astrophyiscs'', Cambridge University Press, Cambridge

\reference{} Gallo, E., Corbel, S., Fender, R. P., Maccarone, T. J.,
\& Tzioumis, A. K., 2004, MNRAS, 347, L52

\reference{} Gallo, E., et al., 2009, in preparation

\reference{} Gammie, C. F., Shapiro, S. L., \& McKinney, J. C., 2004,
ApJ, 602, 312

\reference{} Garcia, M. R., Miller, J. M., McClintock, J. E., King, A. R., \& Orosz, J., 2003, ApJ, 591, 388

\reference{} Gilfanov, M., Churazov, E., \& Revnivtsev, M., 2000, MNRAS, 316, 923

\reference{} Greiner, J., Cuby, J. G., \& McCaughrean, M. J., 2001,
Nature, 414, 522

\reference{} Hannikainen, D., et al., 2001, ApSSS, 276, 45

\reference{} Heger, A., \& Woosley, S., 2002, ApJ, 567, 532

\reference{} Herrero, A., et al., 1995, A\&A, 297, 556

\reference{} Hjellming, R. M., \& Rupen, M. P., 1995, Nature, 375, 464

\reference{} Homan, J., et al., 2001, ApJS, 132, 377

\reference{} Hynes, R. I., Steeghs, D., Casares, J., Charles, P. A.,
\& O'Brien, K., 2003, ApJ, 583, L95

\reference{} Hynes, R. I., Charles, P. A., van Zyl, L., Barnes, A.,
Steeghs, D., O'Brien, K., \& Casares, J., 2004, MNRAS, 348, 100

\reference{} van der Klis, M., 2001, ApJ, 561, 943

\reference{} van der Klis, M., 2006, in ``Compact Stellar X-ray
Sources'', eds. W. Lewin and M. van der Klis, Cambridge University
Press: Cambridge

\reference{} Krolik, J. H., Hawley, J., \& Hirose, S., 2005, ApJ, 622, 1008

\reference{} Laor, A., 1991, ApJ, 376, L90

\reference{} Li, L., Zimmerman, E. R., Narayan, R., McClintock, J. E., 2005, ApJS, 157, 335
\reference{} Maccarone, T., 2002, MNRAS, 336, 1371

\reference{} Martin, R. G., Reis, R. C., \& Pringle, J. E., 2008, MNRAS, in press

\reference{} Martocchia, A., Matt, G., Karas, V., Belloni, T., \&
Feroci, M., 2002, A\&A, 387, 215

\reference{} McClintock, J. E., et al., 2006, ApJ, 652, 518

\reference{} Miller, J. M., et al., 2002, ApJ, 570, L69

\reference{} Miller, J. M., et al., 2002b, ApJ, 577, L15

\reference{} Miller, J. M., et al., 2004a, ApJ, 606, L131

\reference{} Miller, J. M., Fabian, A. C., Nowak, M. A., \& Lewin,
W. H. G. L., 2005, in the proceedings of the Tenth Marcel Grossmann
Meeting, eds. M. Novello, S. Perez Berliaffa, R. Ruffini, Singapore:
World Scientific
 
\reference{} Miller, J. M., Homan, J., Steeghs, D., Rupen, M.,
Hunstead, R. W., Wijnands, R., Charles, P., \& Fabian, A. C., 2006,
ApJ, 653, 525

\reference{} Miller, J. M., et al., 2006, ApJ, 653, 525

\reference{} Miller, J. M., et al., 2006b, Nature, 441, 953

\reference{} Miller, J. M., 2007, ARA\&A, 45, 441

\reference{} Miller, J. M., et al., 2008, ApJ, 679, L113

\reference{} Miller, J. M., et al., 2008b, ApJ, 680, 1359

\reference{} Miniutti, G., Fabian, A. C., \& Miller, J. M., 2004, MNRAS, 351, 466

\reference{} Miniutti, G., \& Fabian, A. C., 2004, MNRAS, 349, 1435

\reference{} Miniutti, G., \& Fabian, A. C., 2009, to appear in ``Kerr
Spacetime: Rotating Black Holes in General Relativity'',
eds. D. L. Wiltshire, M. Visser, \& S. M. Scott, Cambridge University
Press, Cambridge

\reference{} Miniutti, G., et al., 2007, PASJ, 59, 315

\reference{} Mirabel, F., \& Rodrigues, I., 2003, Science, 300, 1119

\reference{} Mitsuda, K., et al., 1984, PASJ, 36, 741

\reference{} Munoz-Darias, T., Casares, J., \& Martinez-Pais, I. G., 2008, MNRAS, 385, 2205

\reference{} Orosz, J., \& Bailyn, C. D., 1997, ApJ, 477, 876

\reference{} Orosz, J., et al., 2002, ApJ, 568, 845

\reference{} Orosz, J., et al., 2004, ApJ, 616 376

\reference{} Park, S. Q., et al., 2004, ApJ, 610, 378

\reference{} Reis, R. C., Fabian, A. C., Miniutti, G., Miller, J. M.,
\& Reynolds, C., 2008, MNRAS, 387, 1489

\reference{} Remillard, R. A., \& McClintock, J. E., 2006, ARA\&A, 44, 49

\reference{} Revnivtsev, M., Sunyaev, R., Gilfanov, M., \& Churazov, E., 2002, A\&A, 385, 904

\reference{} Reynolds, C. S., \& Fabian, A. C., 2008, ApJ, in press,
arxiv:0711.4158

\reference{} Reynolds, C. S., \& Miller, M. C., 2008, ApJ, subm., arxiv:0805.2950

\reference{} Ross, R. R., \& Fabian, A. C., 2007, MNRAS, 381, 1697

\reference{} Rossi, S., Homan, J., Miller, J. M., \& Belloni, T.,
2005, MNRAS, 360, 763

\reference{} Rykoff, E. S., Miller, J. M., Steeghs, D., \& Torres,
M. A. P., 2007, ApJ, 666, 1129

\reference{} Shafee, R., et al., 2006, ApJ, 636, L113

\reference{} Shafee, R., et al., 2008, ApJ, subm.

\reference{} Shibata, M., \& Shapiro, S. L., 2002, ApJ, 572, L39

\reference{} Sikora, M., Stawarz, L., \& Lasota, J. P., 2007, ApJ, 658, 815

\reference{} Sobczak, G. J., et al., 1999, ApJ, 520, 776

\reference{} Stirling, A. M., et al., 2001, MNRAS, 327, 1273

\reference{} Swank, J., Smith, E. A., Smith, D. M., \& Markwardt,
C. B., 2006, ATEL 944

\reference{} Takahashi, H., et al., 2008, PASJ, 60, 69

\reference{} Volonteri, M., Madau, P., Quataert, E., \& Rees, M. J.,
2005, ApJ, 620, 69

\reference{} Wijnands, R., \& Miller, J. M., 2002, ApJ, 564, 974

\reference{} in 't Zand, J. J. M., et al., 2000, A\&A, 357, 520

\reference{} in 't Zand, J. J. M., et al., 2002, A\&A, 390, 597

\reference{} in 't Zand, J. J. M., Miller, J. M., Oosterbroek, T., \&
Parmar, A. N., 2002b, A\&A, 394, 553

\reference{} Zdziarski, A. A., Gierlinski, M., Mikolajewska, J.,
Wardzinski, G., Smith, D. M., Harmon, B. A., Kitamoto, S., 2004,
MNRAS, 351, 791

\reference{} Zhang, S. N,. Cui, W., \& Chen, W., 1997, ApJ, 482, L155

\reference{} Zimmerman, E. R., Narayan, R., McClintock, J. E., \&
Miller, J. M., 2005, ApJ, 618, 832

%----------------------------------------------------------------------

\end{references}
\end{document}